\documentclass[aps,pra,twocolumn,10pt,showpacs,superscriptaddress]{revtex4}

\usepackage{amsfonts}
\usepackage{graphicx}
\usepackage{amsmath}
\usepackage{amsthm}
\usepackage[colorlinks=true, citecolor=blue, urlcolor=blue ]{hyperref}
\usepackage{graphicx}

\begin{document}
\title{Role of complementary correlations in the evolution of classical and quantum correlations under Markovian decoherence}

\author{Prasenjit Deb}
\email{devprasen@gmail.com}
\affiliation{Department of Physics and Center for Astroparticle Physics and Space Science, Bose Institute, Bidhan Nagar
Kolkata - 700091, India.}

\author{Manik Banik}
\email{manik11ju@gmail.com}
\affiliation{Physics and Applied Mathematics Unit, Indian Statistical Institute, 203 B. T. Road, Kolkata 700108, India.}

\begin{abstract}
Quantum correlation lies at the very heart of almost all the non-classical phenomena exhibited by quantum systems
composed of more than one subsystem. In the recent days it has been pointed out that there exists quantum correlation,
namely discord which is more general than entanglement. Some authors have investigated that for certain initial states
the quantum correlations as well as classical correlation exhibit sudden change under simple Markovian noise. We show
that, this dynamical behavior of the both types of correlations can be explained using the idea of \emph{complementary
correlations} introduced in [\href{http://arxiv.org/abs/1408.6851}{arXiv:1408.6851}]. We also show that though certain class 
of mixed entangled states can resist the monotonic decay of quantum correlations,it is not true for all mixed states.
Moreover, pure entangled states of two qubits will never exhibit such sudden change.

\end{abstract}
\maketitle

\section{Introduction}
Quantum information processing protocols \cite{Tele,Dense} require resources which are quantum in nature. In the last few
years it has been proved that different kind of non-classical correlations are necessary resources for performing various
quantum-information-processing tasks. One of the most important non-classical correlations is \emph{quantum entanglement}
\cite{Schrodinger,Werner,Horodecki}, which is a central area of research in quantum information science for a long period
of time. Various information theoretic tasks such as \emph{quantum teleportation} \cite{Tele}, \emph{quantum dense
coding} \cite{Dense}, \emph{quantum key distribution}, \emph{state merging} \cite{SM} can be performed in presence of
entanglement. However, there exists another kind of quantum correlation called \emph{discord} \cite{Zurek} which is more
general than quantum entanglement. Unlike entanglement, quantum discord has received attention during the last few years
and it has been proved to be an useful non classical resource. Researchers have shown that discord provides speedup in
performing some tasks in a non-universal model of quantum computation \cite{Knill}. Some other operational significance
of discord have been also pointed out by others \cite{Streltsov,Adhikari,Chuan,Cubitt,Dagmar,Datta}. On the other hand,
quantum mutual information is the information-theoretic measure of the total correlation in a bipartite quantum state.
Groisman \cite{Groisman}, Schumacher and Westermoreland \cite{Schumacher} showed the significance of quantum mutual
information, which can be thought as the sum of quantum and classical correlation \cite{Vedral}.
In most of the ideal cases it is assumed that quantum systems are isolated from environment and one can use the unitary
evolution to  illustrate the dynamics of such systems. Unfortunately, during the practical applications the quantum
systems interact with environment resulting in loss of quantum coherence which in turn destroys the quantum correlations.
This destruction of quantum properties by the inevitable interaction with the environment is perhaps the major hindrance
to the development of quantum technologies till date. Recently several studies revealed the dynamics of quantum and
classical correlations under both Markovian \cite{Maziero1,Maziero2,Mazzola,Sarandy} and non-Markovian \cite{Fanchini}
decoherence. Interestingly, contrary to the entanglement dynamics where \emph{sudden death} may occur \cite{Yu1,Yu2},
quantum correlation measured by quantum discord dose not exhibit this behavior. However sudden change may occur in the
decoherence process. More specifically, for certain class of states e.g Bell diagonal states, discord remains constant
for a particular period of time and then decays, while classical correlation decays first and then becomes constant
\cite{Mazzola}.
  
In this paper we have focused  mainly on two questions:
\begin{enumerate}
\item[(1)]  What are the underlying  physical mechanisms for which  classical and quantum correlations (measured in terms
of discord) suffer sudden change in the decoherence process as studied by Mazzola \emph{et. al} \cite{Mazzola}?
\item[(2)] Does mixedness provide some advantage to prevent the loss of quantum coherence, and hence quantum
correlations?
\end{enumerate}

To answer the $1^{st}$ question we have used the idea of \emph{complementary correlations} introduced in \cite{Lorenzo,Wu}
and we have succeed to describe the physical mechanism going under the phenomena observed in \cite{Mazzola}. More
precisely, we have shown that under some restrictions for certain class of bipartite qubit states the amount of classical
correlation and quantum discord are exactly equal to the correlations between the complementary observables of two sides.
We have also found out the initial class of states exhibiting such behavior. With a lot of surprise, we have found that
the two qubit pure entangled states will never show the said behavior. Rather, some particular  mixture of pure states
exhibit sudden change in the classical and quantum correlations in the decoherence process which answers the $2^{nd}$
question affirmatively.

The organization of the paper is as follows: Section (\ref{sec2}) contains a brief overview on quantum discord and mutual
infomation; in section (\ref{sec3}) we give a brief description on complementary correlations and discuss about
correlation tensor matrix; Markovian decoherence and quantum channels are discussed briefly in section (\ref{sec4}); we
present our results in section (\ref{sec5}) and Section (\ref{sec6}) contains discussions and concluding remarks.

\section{Discord and Mutual Information}\label{sec2}

Entanglement is perhaps the most familiar non-classical correlation observed in quantum systems composed with more than
one subsystem. But, it is important to note that some states with zero entanglement can perform tasks which are not
possible in the classical regime. It is due to the fact that those states have nonclassical correlation even though they
are unentangled. Apart from entanglement, the measure of quantum correlation that has received a great deal of attention
is \emph{Quantum Discord} ($\mathcal{D}$), originally proposed by Ollivier and Zurek \cite{Zurek}. The quantum discord is
defined as: 
\begin{equation}
\mathcal{D}(\rho_{AB})\equiv \mathcal{I}(\rho_{AB}) - \mathcal{C}(\rho_{AB}),\label{discord}
\end{equation}
where, $\mathcal{C}(\rho_{AB})$ and  $\mathcal{I}(\rho_{AB})$ are respectively the classical correlations and quantum 
mutual information of the bipartite state $\rho_{AB}$. Quantum  mutual information $\mathcal{I}(\rho_{AB})$ measures the
total correlation present in the state $\rho_{AB}$, and it is defined as: 
\begin{equation}
\mathcal{I}(\rho_{AB})= S(\rho_{A}) + S(\rho_{B}) - S(\rho_{AB}),
\end{equation}
where, $\rho_{A}$ and $\rho_{B}$ are the reduced density matrix of the subsystems $A$ and $B$ respectively and  $S(\rho)=
-$Tr$ \{\rho \log_{2}\rho\}$ is the von Neumann entropy. From the above definition of mutual information it is clear that
one can straightforwardly calculate it for a given state $\rho_{AB}$. The maximum value of $\mathcal{I}(\rho_{AB})$ is
$2\log_2d$, achieved by two-qudit maximally entangled states and it's minimum value is zero, achieved by product states. 

On the other hand, classical correlations, $\mathcal{C}(\rho_{AB})$ of a composite quantum state can be quantified via
the measure proposed by Henderson and Vedral \cite{Vedral} which is given by:
\begin{equation}
\mathcal{C}(\rho_{AB})\equiv \max_{\{\Pi_j\}} [S(\rho_A)- S_{\{\Pi_j\}}(\rho_{A\arrowvert B})]\label{vedral},
\end{equation}
where, the maximum is taken over the set of projective measurements $\{\Pi_j\}$ on subsystem $B$.
$S_{\{\Pi_j\}}(\rho_{A\arrowvert B})= \Sigma_{j}  p_j~ S(\rho^j_A)$ is the  entropy of subsystem $A$ conditioned on $B$,
$\rho^j_A = $Tr$_B(\Pi_j \rho_{AB} \Pi_j)/p_j$ is the density matrix of subsystem $A$ depending on the measurement
outcome of $B$ and $p_j= \mbox{Tr}_{AB}(\rho_{AB}\Pi_j)$ is the probability of $j^{th}$ outcome. Note that, unlike
$\mathcal{I}(\rho_{AB})$, the classical correlation is asymmetric with respect to the subsystems involved and so is
$\mathcal{D}(\rho_{AB})$.  

As the definition of classical correlation requires optimization over all possible projective measurements (more
generally over the positive operator valued measurements \cite{chuang}) that can be performed at one part of the
composite system, it is in general very hard to find the amount of quantum discord for arbitrary bipartite state. To
avoid this complexity different computably easy measures of quantum discord have been proposed recently
\cite{Tufarelli,Fu,Dakic}. Few important properties of quantum discord are the following:
\begin{itemize}
\item [(a)]  $\mathcal{D}(\rho_{AB})\geq 0  ,~~~~ \forall ~~~  \rho_{AB}$.\\ 
The proof is straight forward. Putting the explicit form of $\mathcal{I}(\rho_{AB})$ and $\mathcal{C}(\rho_{AB})$ in Eq.(\ref{discord}), we get
\begin{equation}
\mathcal{D}(\rho_{AB})= S_{\{\Pi_j\}}(\rho_{A\arrowvert B}) - S(\rho_{A\arrowvert B}),
\end{equation}
where, $S(\rho_{A\arrowvert B})= S(\rho_{AB}) - S(\rho_B)$ is the quantum conditional entropy. Finding the value of 
$S_{\{\Pi_j\}}(\rho_{A\arrowvert B})$ involves measurement on B side which typically increases the entropy and hence
$S_{\{\Pi_j\}}(\rho_{A\arrowvert B})\geq S(\rho_{AB}) - S(\rho_B)$, which completes the proof. If we put our concern on  
quantum conditional density matrix $\rho_{A\arrowvert B}$, then from ref. \cite{Cerf} it can be said that this 
conditional density matrix retains the quantum phases and coherence. So physically quantum discord is a measure
of information which cannot be extracted without joint measurements \cite{Zurek}.
\item[(b)] $\mathcal{D}(\rho_{AB})=0 $ for quantum-classical (QC) states which are of the form: 
\begin{equation}
\rho_{QC}= \Sigma_i~p_i \rho^i_A\otimes\lvert i_B\rangle\langle i_B\lvert,\label{cc}
\end{equation}
where $\{\lvert i_B\rangle\}$ is an orthonormal basis for subsystem $B$, and $\rho_A^i$'s are density matrices of subsystem $A$ and 
$\{p_i\}$ is a probability distribution.
For two-qudit maximally entangled states $\mathcal{D}(\rho_{AB})=\log_2d$. 

\item[(c)] Quantum discord is non-increasing under completely positive trace preserving (CPTP) maps on unmeasured party $A$
\cite{adesso}, i.e,
\begin{equation}
\mathcal{D}(\rho_{AB})\geq \mathcal{D}([\Lambda_A\otimes I_{B}]\rho_{AB}),
\end{equation}
where, $\Lambda_{A}$ being the CPTP map on $A$ \cite{choi}. 
\end{itemize}

\section{Complementary correlations and correlation tensor matrix}\label{sec3}
{\bf Complementary correlations}: The concept of complementary correlations is recently introduced in ref.\cite{Lorenzo}. 
Consider a quantum mechanical system described by $d$-dimensional Hilbert space. 
Let $\mathcal{M}$ and $\mathcal{N}$ are two observables acting on the system with $\{|m_i\rangle\}_{i=1}^d$ and $\{|n_j\rangle\}_{j=1}^d$ denoting the non-degenerate
eigenstates, respectively. $\mathcal{M}$ and $\mathcal{N}$ are called \emph{complementary observables}, if $|\langle m_i|n_j\rangle|^2 =\frac{1}{d}$, for all $i,j$.
It means that if one knows the value of one of the complementary observables i.e., if the system is prepared
in one of the eigenstates of one of the complementary observables then all the possible values of the other observable are equal probable. 

The authors of \cite{Lorenzo} have shown that, for a composite quantum system the 
correlations in the measurement of such complementary observables is a good signature of quantum correlations
present in the state. If two quantum systems of finite dimension are considered and two observables $\mathcal{A}\otimes\mathcal{B}$ and 
$\mathcal{C}\otimes\mathcal{D}$ are taken into account, where $\mathcal{A}$ and $\mathcal{C}$ are complementary observables on one 
subsystem and $\mathcal{B}$ and $\mathcal{D}$ 
on the other, then the quantity $|\chi _{\mathcal{AB}}|+|\chi_{\mathcal{CD}}|$ denotes the value of 
\emph{complementary correlations} with $|\chi _{\mathcal{AB}}|$ and $|\chi_{\mathcal{CD}}|$ denoting the absolute value of
correlations on complementary observables. The sum not only tells about the quantum correlations present in a composite
quantum system but it also represents the overall correlations of the composite system.

\begin{figure}[t!]
\centering
\includegraphics[height=5cm,width=8cm]{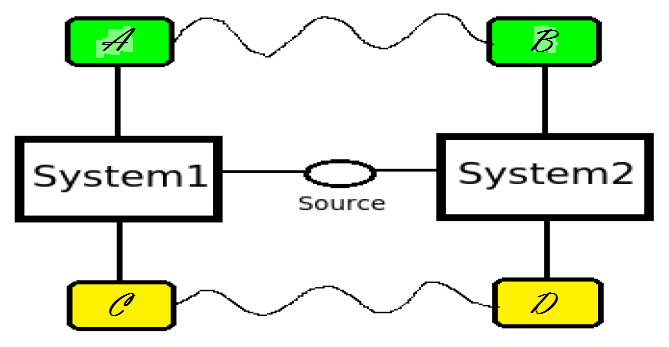}
\caption{(Color on-line) System 1 and System 2 are the subsystems of some composite quantum system. $\mathcal{A}$ and $\mathcal{C}$ are 
the complementary observables for system 1, while $\mathcal{B}$ and $\mathcal{D}$ are the same complementary observables for system 2
}\label{fig1}
\end{figure}
Consider that $\chi_{\mathcal{AB}} =\mathcal{I}_{\mathcal{AB}}$ and $\chi_{\mathcal{AB}} = \mathcal{I}_{\mathcal{CD}}$, where $\mathcal{I}$ 
is the mutual information defined earlier and having an alternate definition as:
\begin{equation}
\mathcal{I}_{\mathcal{AB}}\equiv H(\mathcal{A})-H(\mathcal{A}|\mathcal{B})\label{mutual},
\end{equation}
where, $H(\mathcal{A})$ is the Shannon entropy of the outcome probabilities of the measurement $\mathcal{A}$ performed on the first system 
and $H(\mathcal{A}|\mathcal{B})$ denotes 
the conditional entropy, conditioning being done on second system. Therefore in terms of mutual information the complementary 
correlations reads as: $\mathcal{I}_{\mathcal{AB}}+\mathcal{I}_{\mathcal{CD}}$. It can be easily shown that:
\begin{enumerate}
\item[(i)] If $\mathcal{I}_{\mathcal{AB}}+\mathcal{I}_{\mathcal{CD}} = 2\log_2d$, then the bipartite quantum system is 
maximally entangled and there exists two complementary measurement bases or in other words mutually unbiased bases (MUBs).
\item[(ii)] If $\mathcal{I}_{\mathcal{AB}}+\mathcal{I}_{\mathcal{CD}}>\log_2d$, then there is entanglement in the composite system.
\item[(iii)] If $\mathcal{I}_{\mathcal{AB}}+\mathcal{I}_{\mathcal{CD}}=\log_2d$, the bipartite state is a classically correlated (CC) state which 
belongs to the set of separable states and the quantum correlations for such states is zero.
\end{enumerate}
 
{\bf Correlation tensor matrix}: Here we concentrate on two-qubit quantum system defined on the Hilbert space $\mathcal{H}= C^2 \otimes
C^2 $. The collection of Hermitian operators acting on $\mathcal{H}$ constitute a inner product space with
Hilbert-Schmidt inner product defined as $\langle\alpha,\beta\rangle = $Tr$ (\alpha^\dagger\beta)$, where $\alpha$ and $\beta$ are Hermitian operators
acting on $\mathcal{H}$. In such a Hilbert-Schmidt space any generic state of the system can be expressed as \cite{Horodecki1}:
\begin{equation}
\rho=\frac{1}{4}(\mathbf{1}_2\otimes\mathbf{1}_2+\mathbf{a}\cdot\boldsymbol{\sigma}\otimes\mathbf{1}_2 +\mathbf{1}_2\otimes \mathbf{b}\cdot\boldsymbol{\sigma} +\sum _{m,n=1}^{3}c_{nm}\sigma_n\otimes\sigma_m )\label{2qubit}
\end{equation}
where, $\mathbf{1}_2$ is the identity operator, $\mathbf{a}$ and $\mathbf{b}$ denote the local Bloch vectors for each subsystem and $\{\sigma_n\}_{n=1}^{3}$ are the standard Pauli spinors $\sigma_x$,
$\sigma_y$ and $\sigma_z$. The $3\times3$ matrix $\mathcal{T}$ formed by the coefficients $c_{nm}$ is called the correlation tensor matrix
as it is responsible for the correlations:
\begin{equation}
\mathcal{E}(\mathbf{a,b})\equiv\mbox{Tr}(\rho\mathbf{a}\cdot\boldsymbol{\sigma}\otimes \mathbf{b}\cdot\boldsymbol{\sigma})=(\mathbf{a},\mathcal{T}\mathbf{b}).
\end{equation}
Note that $c_{nm} =\mbox{Tr}(\rho\sigma_n\otimes\sigma_m)$ are the \emph{expectation values}  of the 
observables $\sigma_n\otimes\sigma_m$.  The state $\rho$, as expressed in Eq.(\ref{2qubit}) can always be transformed to a state 
$\widetilde{\rho}$ for which  the matrix $\mathcal{T}$ becomes diagonal by acting local unitaries $U_1$ and $U_2$. Though the unitaries
transform the state $\rho$ to $\tilde{\rho}$, the inseparability (separability) remains invariant. The transformed state can be represented as 
\begin{equation}
\widetilde{\rho}= U_1\otimes U_2\rho U_1^{\dagger}\otimes U_2^{\dagger}.
\end{equation}

The transformation of the state $\rho$ is possible due to the fact that for any unitary transformation $U$ there  is always
a unique rotation $O$ such that $U\mathbf{\hat n}\cdot\boldsymbol{ \sigma} U^{\dagger} = (O\bf{\hat n})\cdot \boldsymbol{\sigma }$ 
and the parameters $\bf a$, $\bf b$ and $\mathcal{T}$ transform themselves as ${\bf {a^\prime}} = O_1\bf a$,
${\bf {b ^ \prime}}=O_1\bf b$ and $\mathcal{T}^\prime = O_1\mathcal{T}O_2^{\dagger}$ respectively, where $ \bf{a^{\prime}}$,
$\bf{b^{\prime}}$ and $\mathcal{T}^{\prime}$ are the new parameters for the state $\widetilde{\rho}$. 

As the unitaries $U_1$ and $U_2$ diagonalize the correlation tensor matrix, hence we have,
\begin{equation}
\mathcal{T}^{\prime} = \begin{pmatrix} c_{1}&0&0 \\ 0&c_{2}&0 \\0&0&c_{3} \end{pmatrix},
\end{equation}
where, $c_1=\mbox{Tr}(\rho \sigma _x \otimes \sigma _x)$, $c_2=\mbox{Tr}(\rho \sigma _y \otimes \sigma _y)$, 
$c_3 =\mbox{Tr}( \rho \sigma _z \otimes \sigma _z)$, are the expectation values of the observables $\sigma _x\otimes\sigma _x $,
$\sigma _y\otimes\sigma _y $ and  $\sigma _z \otimes \sigma _z $, respectively.

\section{Markovian decoherence and quantum channels}\label{sec4}

\emph{\bf Markovian decoherence}:
Decoherence is a physical process which describe the gradual loss of coherence present in any quantum system \cite{zurek2}. 
The process can be represented by some family of linear maps
$\{\Lambda_{(t_2,t_1)}, t_2\geq t_1 \geq t_0\}$, where, $t_0$, $t_1$ and $t_2$ denote time \cite{wolf}. If the linear map
$\Lambda_{(t_2,t_1)}$ satisfies following three properties i.e.,
\begin{enumerate}
\item[(i)] trace-preserving i.e., $\mbox{Tr}(\rho)=\mbox{Tr}(\Lambda[\rho])$,
\item[(ii)] completely positive i.e., $\Lambda\otimes\mathbf{1}_{C^n}$ is positive for all $n$, where $\mathbf{1}_{C^n}$ denote identity map acting 
on n-dimensional Hilbert space and 
\item[(iii)] $\Lambda_{(t_3,t_1)} =  \Lambda_{(t_3,t_2)}$  $\Lambda_{(t_2,t_1)}$, 
\end{enumerate}
then the decoherence process is called 
\emph{Markovian} decoherence \cite{wolf}. The linear map $\Lambda(.)$ basically describes the time evolution of 
the quantum system that interacts with its environment. The above mentioned conditions, which a linear map should follow to represent a 
Markovian process comes from a very useful theorem, namely the Kraus representation theorem \cite{kraus} which provides the following
operator sum representation for any CPTP map,
\begin{equation}
\Lambda(A) = \sum_{j=1}^{r} \Gamma_j A\Gamma_j^{\dagger}
\end{equation}
where, $\Gamma_j$ are the \emph{Kraus operators} and $r$ is the Kraus rank which determines the number of Kraus operators 
in the operator - sum representation of the linear map $\Lambda(.)$. The normalization principle leads to the fact that
$\Lambda(.)$ is trace preserving \emph{iff} $\sum_{j} \Gamma_j^{\dagger}\Gamma_j = 1 $.

\emph{\bf Quantum Channels}:
In any  communication protocol one has to send information through  channels. If the information is quantum in nature, then
the time evolution of the quantum system carrying the information can be modelled by quantum channels. 
Mathematically, quantum channels are some \emph{superoperators} or CPTP linear maps having the operator-sum representation.
The three important classes of quantum channels are \emph{depolarizing
channel, amplitude damping channel} and \emph{phase damping channel}.
\begin{itemize}
\item[(I)] {\bf Depolarizing channel}: The Kraus operators which represent  the depolarizing channel are:
$$\Gamma_0 = \sqrt{1-p}~{I},~~~~\Gamma_1 = \sqrt{\frac{p}{3}}{\sigma_1},$$
$$\Gamma_2 = \sqrt{\frac{p}{3}}{\sigma_2},~~~~~~\Gamma_3 = \sqrt{\frac{p}{3}}{\sigma_3},$$
where, $p$ represents probability. Under depolarizing channel any general density matrix $\rho$ evolves as: 
$$\rho \rightarrow \rho^{\prime} = (1-p)\rho+ \frac{p}{3}(\sigma_1\rho\sigma_1 $$
\vspace{-.7cm} 
$$~~~~~~~~~~~~~~~~~~~~~~~~~~~~~~~~+ \sigma_2\rho\sigma_2 + \sigma_3\rho\sigma_3).$$
\item[(II)] {\bf Amplitude damping channel}: Kraus operators representing amplitude damping channel as a schematic model
of the decay of an excited state  of a two state quantum system are as follows:
$$ \Gamma_0 = \begin{pmatrix} 1 & 0 \\ 0& \sqrt{1-p} \end{pmatrix},~~ 
\Gamma_1 = \begin{pmatrix} 0&\sqrt{p} \\ 0&0 \end{pmatrix}.$$
The evolution a density matrix $\rho$ is given by:
$$\rho \rightarrow \Lambda(\rho) = \Gamma_0 \rho \Gamma_0^{\dagger} +  \Gamma_1 \rho \Gamma_1^{\dagger}.$$

\item[(III)] {\bf Phase damping channel}: Caricaturing of decoherence process in realistic physical situations is possible
by phase damping channel. Though there are phenomenological models leading to decoherence, but none of the models  represent 
the real physical situations. Nevertheless, using operator-sum representation the evolution of a quantum state under
phase damping channel can easily be understood.
The Kraus operators required to represent phase-damping channel are 
$$\Gamma_0 = \sqrt{1-\frac{p}{2}}~ {I}, 
\Gamma_1 = \sqrt{\frac{p}{2}}~ \sigma_3.$$

The most general single-qubit density matrix can be written as: 
\begin{equation*}
\rho =  \begin{pmatrix} \rho_{00}&\rho_{01} \\ \rho_{10}&\rho_{11} \end{pmatrix}
\end{equation*}
where, the diagonal real elements represent the probabilities of finding the qubit  in the state $|0\rangle$ or
$|1\rangle$, respectively, if measurement is done in $\sigma_z$ basis. The off-diagonal elements (\emph{quantum coherences})
have no classical analogue and the phase-damping channel induces a decay of those elements resulting in decoherence.
The single-qubit density matrix evolves as
\begin{eqnarray}
\Lambda(\rho) &=& \Gamma_0\rho\Gamma_0^{\dagger} + \Gamma_1\rho\Gamma_1^{\dagger} \nonumber\\
&=& (1-\frac{p}{2})\rho + \frac{p}{2}\rho \nonumber\\
&=& \begin{pmatrix} \rho_{00}& \rho_{01}(1-p)\\ \rho_{01}(1-p)&\rho_{11} \end{pmatrix}.
\end{eqnarray}
It is clear from the above equation that the off-diagonal terms will gradually decay as the time elapse and the initial coherent
superposition will turn into incoherent superposition or mixture, i.e, 
$$\rho \rightarrow \rho^{\prime} = \lvert \rho_{00}\lvert^2~ \lvert 0\rangle\langle0\lvert
 +\lvert\rho_{11}\lvert^{2}~\lvert1\rangle\langle1\lvert.$$
The phase-damping channel plays the central role in the transition from the quantum to the classical world.
The decay of the off-diagonal terms and hence decoherence can be well understood if we consider the interaction
of the qubit with the environment as a rotation (\emph{phase-kick}) about $z$-axis of the Bloch sphere through an angle $\theta$,
due to which the axes transform as $x^{\prime}=e^{-\lambda}x$, $y^{\prime}=e^{-\lambda}y$ and $z^{\prime}=z$, where $\lambda$ is the damping parameter. In other 
words it can be said that the channel picks out a prefered basis for the qubit, which is $\{\lvert0\rangle , \lvert 1\rangle\}$,
as $z$-basis is the only one in which bit flip never occurs.
\end{itemize}

\section{Results}\label{sec5}
\subsection{Complementary correlations and decoherence}
Consider a generic two qubit bipartite state $\rho_{AB}$ as expressed in Eq.(\ref{2qubit}). 
Applying local unitaries let us diagonalize the correlation tensor matrix and transform the state 
$\rho_{AB}$ to $\rho_{AB}^{\prime}$ so that
\begin{equation}
\rho_{AB}^{\prime} = \frac{1}{4}(\mathbf{1}_2\otimes\mathbf{1}_2+ \mathbf{a}\cdot\boldsymbol{\sigma}\otimes\mathbf{1}_2+\mathbf{1}_2\otimes \mathbf{b}\cdot\boldsymbol{\sigma}+\sum _{n=1}^{3} c_{n}\sigma_n\otimes\sigma_n ).
\end{equation}
For simplicity we consider the states with maximally mixed marginals, i.e, $\bf a=0$ and $\bf b=0$. Thus we have:
\begin{equation} 
\rho_{AB}^{\prime} = \frac{1}{4}(\mathbf{1}_2\otimes\mathbf{1}_2+\sum _{n=1}^{3}c_{n}\sigma_n\otimes\sigma_n),\label{bell-diag}
\end{equation}
here, $c_n$'s are the diagonal elements of correlation tensor matrix $\mathcal{T}$ and $0\leq|c_n|\leq 1$. The class 
of states represented by $\rho_{AB}^{\prime}$ in Eq.(\ref{bell-diag}) are called the Bell diagonal states, which includes pure
Bell states ($\lvert c_1\lvert = \lvert c_2\lvert= \lvert c_3 \lvert =1$) and Werner class of
states ($\lvert c_1\lvert = \lvert c_2\lvert= \lvert c_3 \lvert=\mbox{c}$) \cite{Mazzola}.

When both the subsystems of the composite state of Eq.(\ref{bell-diag}) are subjected to local Markovian noise, the time evolution of the composite state is given by:
\begin{eqnarray}
\rho_{AB}^{\prime}(t)&=&\lambda_{\Psi}^{+}(t)\lvert\Psi^{+}\rangle\langle\Psi^{+}\lvert + \lambda_{\Phi}^{+}(t)\lvert\Phi^{+}\rangle\langle\Phi^{+}\lvert \nonumber\\
&&+ \lambda_{\Phi}^{-}(t)\lvert\Phi^{-}\rangle\langle\Phi^{-}\lvert + \lambda_{\Psi}^{-}(t)\lvert\Psi^{-}\rangle\langle\Psi^{-}\lvert,
\end{eqnarray}
where, 
\begin{eqnarray}
\lambda_{\Psi}^{\pm}(t)&=&\frac{1}{4}[1\pm c_1(t) \mp c_2(t) + c_3(t)],\label{psi-pm}\\
\lambda_{\Phi}^{\pm}(t)&=&\frac{1}{4}[1\pm c_1(t) \pm c_2(t) - c_3(t)],\label{phi-pm}
\end{eqnarray}
 and $\lvert\Phi^{\pm}\rangle = \frac{1}{\sqrt{2}}(\lvert 00\rangle \pm \lvert 11 \rangle) , 
\lvert\Psi^{\pm}\rangle = \frac{1}{\sqrt{2}}(\lvert 01\rangle \pm \lvert 10 \rangle)$ are Bell states. 

If we  consider
phase damping channel as the local Markovian noise, then the co-efficients in Eq.(\ref{psi-pm}) and Eq.(\ref{phi-pm}) will be:
\begin{eqnarray}
c_1(t) = c_1(0)e^{-2\gamma t},\nonumber\\
c_2(t) = c_2(0)e^{-2\gamma t},\nonumber\\
c_3(t) = c_3(0)\equiv c_3,\label{time-evo}
\end{eqnarray}
where $\gamma$ is the phase damping rate.
For our analysis we consider the initial states as $c_1(0)= \pm 1$ and $c_2(0)=\mp c_3(0)$ with the condition $\lvert c_3\lvert \leq 1$ , as
considered by other authors \cite{Mazzola} also.
Thus the states read as:
\begin{equation}
\rho_{AB}=\frac{(1+c_3)}{2} \lvert\Psi^{\pm}\rangle\langle\Psi^{\pm}\lvert +\frac{(1-c_3)}{2} \lvert\Phi^{\pm}\rangle\langle\Phi^{\pm}\lvert.\label{an-state}
\end{equation}
The subsystems of the above state are qubit and $\sigma_x$ and  $\sigma_z$ are \emph{complementary} observables for a qubit quantum system. 
From the definition of complementary correlations, the total complementary correlations is therefore:
\begin{equation}
\mathcal{I}^c[\rho_{AB} (t)] = \mathcal{I}(\sigma_ x ^A :\sigma_ x ^B) + \mathcal{I}(\sigma_z^A :\sigma_z^B),\label{mu1}
\end{equation}
where, the superscript ``c'' signifies complementarity. The first term on the right hand side of the Eq.(\ref{mu1}) 
denotes the correlation between the outcomes of $\sigma_ x $ measurement performed on both sides, similarly second term denotes the same, but for $\sigma_z$.
 Using Eq.(\ref{mutual}) we have:
\begin{eqnarray}
\mathcal{I}(\sigma_ x ^A :\sigma_ x ^B)&=&\mathcal{P}[c_1(t)]+\mathcal{P}[-c_1(t)],\label{mu2}\\
\mathcal{I}(\sigma_ z ^A :\sigma_ z ^B)&=&\mathcal{P}[c_3]+\mathcal{P}[-c_3],\label{mu3}
\end{eqnarray}
where $\mathcal{P}[\alpha]=\frac{1+\alpha}{2}\log_2(1+\alpha)$. Inserting Eq.(\ref{mu2})-(\ref{mu3}) in Eq.(\ref{mu1}) we get: 
\begin{eqnarray}
\mathcal{I}^c[\rho_{AB}(t)]&=&\mathcal{P}[c_1(t)]+\mathcal{P}[-c_1(t)]+\mathcal{P}[c_3]\nonumber\\
&&+\mathcal{P}[-c_3].
\end{eqnarray}
Interestingly, for the concerned class of states $\mathcal{I}^c[\rho_{AB}(t)]$ is exactly equal to the mutual information $(\mathcal{I})$ of the state
$\rho_{AB}(t)$, i.e.,
\begin{equation}
\mathcal{I}^c[\rho_{AB}(t)] = \mathcal{I}(\rho_{AB}(t)).
\end{equation}

On the other hand, the classical correlation $\mathcal{C}(\rho_{AB}(t))$ in this case turns out to be:
\begin{equation}
\mathcal{C}(\rho_{AB}(t))=\mathcal{P}[\mathcal{K}(t)]+\mathcal{P}[-\mathcal{K}(t)],
\end{equation}
where $\mathcal{K}(t)=\max\{\lvert c_1(t)\lvert, \lvert c_2(t) \lvert, \lvert c_3(t) \lvert \}$. It is to be noted that
the coefficients $c_1(t)$, $c_2(t)$ and $c_3(t)$ are the expectation values of $\sigma_1\otimes\sigma_1$,
$\sigma_2\otimes\sigma_2$ and $\sigma_3\otimes\sigma_3$ respectively. So, during the calculation of classical correlations
of the specific states considered, the conditional entropy in the Eq.(\ref{vedral}) reaches the minimum when the projective
measurements are performed on eigenstate of that complementary observable $\sigma_n^{(B)}$ for which
$\mbox{Tr}(\rho_{AB}\sigma_n^{(A)}\otimes\sigma_n^{(B)})$ is maximum. Hence, we conclude that, for the class of states taken into
consideration, $\sigma_x$,$\sigma_y$,$\sigma_z$ form a set of complementary observables and classical correlation 
$\mathcal{C}(\rho_{AB}(t))$ is  
\begin{equation}
\mathcal{C}(\rho_{AB}(t)) = \max_{l\in x,y,z}[\mathcal{I}(\sigma_l^A:\sigma_l^B)].
\end{equation}
For our purpose we assume, 
\begin{equation}
\mathcal{I}(\rho_{AB}) = \mathcal{Q}(\rho_{AB}) + \mathcal{C}(\rho_{AB}),\label{assume}\\
\end{equation}
which in turn yields
\begin{equation}
\mathcal{I}^c[\rho_{AB}(t)] = \mathcal{Q}(\rho_{AB}) + \mathcal{C}(\rho_{AB}).
\end{equation}

We are now in a position to explain the sudden transition in classical and quantum decoherence for the states considered in Eq.(\ref{an-state}).

(i) At time $t=0$, Tr$(\rho_{AB}\sigma_1^{(A)}\otimes\sigma_1^{(B)})=c_1(0)=1$. So projective
measurements on eigen-states of $\sigma_x$ will yield  minimum conditional entropy S$(A\lvert B)$ and hence the amount
of classical correlations, which is equal to the correlation between the measurement outcomes of $\sigma_x$ on both sides
of the bipartite state reads as 
\begin{eqnarray}
\mathcal{C}(\rho_{AB}(t_0)) &=& \mathcal{I}(\sigma_ x ^A : \sigma_ x ^B) \nonumber \\
&=&2\mathcal{P}[c_1(t_0)].
\end{eqnarray}
From Eqs. (\ref{discord}), (\ref{mu3}) and (\ref{assume}), the value of quantum correlations or discord comes out to be
\begin{eqnarray}
\mathcal{D}(\rho_{AB}(t_0)) &=& \mathcal{I}(\sigma_ z ^A : \sigma_ z ^B) \nonumber \\
&=& 2\mathcal{P}[c_3].
\end{eqnarray}

(ii) In the time interval $0<t<t^{\prime}= - \ln (\lvert c_3\lvert)/2\gamma $, $c_1(t)>c_3$. So due to the same reasons as
described above, the classical correlations($\mathcal{C}$) and discord($\mathcal{D}$) of the  initial state  will be
\begin{eqnarray}
\mathcal{C}(\rho_{AB}(t)) &=& \mathcal{I}(\sigma_ x ^A : \sigma_ x ^B) \nonumber \\
&=&\mathcal{P}[c_1(t)]+\mathcal{P}[-c_1(t)],\\
\mathcal{D}(\rho_{AB}(t)) &=& \mathcal{I}(\sigma_ z ^A : \sigma_ z ^B) \nonumber \\
&=&\mathcal{P}[c_3]+\mathcal{P}[-c_3].
\end{eqnarray}
However, in this time interval the classical correlation $\mathcal{C}(\rho_{AB}(t))$ decays and discord $\mathcal{D}(\rho_{AB}(t))$
remains constant. In other words we can say that the correlations $ \mathcal{I}(\sigma_ x ^A :\sigma_ x ^B)$  decay and
correlations $ \mathcal{I}(\sigma_ z ^A :\sigma_ z ^B)$ remain constant. The physical origin of this fact is that 
at $t>0$ the phase-damping channel induces a decay in the quantum coherence(phase) of the state $\rho_{AB}(t)$,
which results a decay in the expectation value $c_1(t)= $Tr$[\rho_{AB}(t)\sigma_1^A\otimes\sigma_1^B]$, whereas
the expectation value $c_3(t)= $Tr$[\rho_{AB}(t)\sigma_3^A\otimes\sigma_3^B]$ remains constant. 
The phase-damping channel picks the $\{\lvert 0\rangle, \lvert 1 \rangle\}$ basis as preferred basis for each qubit and 
destroys all other superpositions of $\lvert 0\rangle$ and $\lvert 1 \rangle$, resulting in the decay of $c_1(t)$ and
hence $\mathcal{C}[\rho_{AB}(t)]$. Here the discord remains constant.

(iii) For $t>t^{\prime}$, $c_1(t)<c_3$. So, in this case the conditional entropy $S(A\lvert B)$ will be minimum if 
projective measurements are performed on eigenstates of $\sigma_z$. Thus the classical correlation will be:
\begin{eqnarray}
\mathcal{C}(\rho_{AB}(t)) &=& \mathcal{I}(\sigma_ z ^A : \sigma_ z ^B) \nonumber \\
&=&\mathcal{P}[c_3]+\mathcal{P}[-c_3].
\end{eqnarray}
From Eq.(\ref{discord}), (\ref{mu2}) and (\ref{assume}), discord of $\rho_{AB}(t)$ will be:
\begin{eqnarray}
\mathcal{D}(\rho_{AB}(t)) &=& \mathcal{I}(\sigma_ x ^A : \sigma_ x ^B) \nonumber \\
&=&\mathcal{P}[c_1(t)]+\mathcal{P}[-c_1(t)].
\end{eqnarray}
Hence, for $t>t^{\prime}$, the classical correlation is constant in time, whereas, discord starts to decay. To understand
this sudden change in the evolution of classical and quantum correlations for $t>t^{\prime}$, we focus on the definition
of classical correlations and notice that during this time projective measurement on the eigenstates of $\sigma_z$ will
yield the classical correlations, i.e., only the term $\mathcal{I}(\sigma_z^A:\sigma_z^B)$ contributes in $\mathcal{C}(\rho_{AB}(t))$. Now, as the channel is phase damping channel, the expectation value $c_3$ does not change with time
and as a result classical correlation becomes constant after time $t^{\prime}$.
\paragraph*{}
On the other hand, during this time the correlation between measurement outcomes of another complementary observable $\sigma_x$
represents the amount of $\mathcal{D}(\rho_{AB}(t))$, present in the state $\rho_{AB}(t)$. Whereas, it is mentioned before 
that the expectation value $t_x$ decays due to the effect of phase daming noise. Therefore, $\mathcal{D}(\rho_{AB}(t))$
decays after time $t^{\prime}$.

\subsection{Mixedness and decoherence}
We now focus on two questions.(i) Will the pure entangled two qubit states exhibit sudden change in the evolution of 
$\mathcal{C}(\rho_{AB}(t))$ and $\mathcal{D}(\rho_{AB}(t))$ ? (ii) Does mixedness will always ensure sudden change
in decoherence process? To answer the first question, we consider a pure two qubit entangled state
\begin{equation}
\lvert\Psi_{AB}\rangle = a\lvert 00\rangle + b \lvert 11\rangle 
\end{equation}
where, ${\lvert a\lvert}^2$ and ${\lvert b\lvert}^2$ are probabilities. This state belongs to the class of states for which
the correlation tensor matrix $(\mathcal{T})$ is diagonal. The three diagonal elements or the  expectation values of the 
observables $\sigma_1\otimes\sigma_1$, $\sigma_2\otimes\sigma_2$ and  $\sigma_3\otimes\sigma_3$ are found to be,
\begin{eqnarray}
c_1= ({\lvert a \vert}^2 + {\lvert b \vert}^2)=1,\nonumber\\
c_2= -({\lvert a \vert}^2 + {\lvert b \vert}^2)=-1,\nonumber\\
c_3= ({\lvert a \vert}^2 + {\lvert b \vert}^2)=1.
\end{eqnarray}
If both the qubits of the state are subjected to local phase damping channel, then all the $c_i$'s change as
Eq.(\ref{time-evo}). The classical correlation $\mathcal{C}(\rho_{AB}(t))$ and discord $\mathcal{D}(\rho_{AB}(T))$
of the state for $t>0$ read as:
\begin{eqnarray}
\mathcal{C}(\rho_{AB}(t)) &=& \mathcal{I}(\sigma_ z ^A : \sigma_ z ^B) \nonumber \\
&=&\mathcal{P}[c_3]+\mathcal{P}[-c_3]= 1,\label{cla-core}\\
\mathcal{D}(\rho_{AB}(t))&=&\mathcal{I}(\sigma_ x ^A :\sigma_ x ^B)\nonumber \\
&=&\mathcal{P}[c_1(t)]+\mathcal{P}[-c_1(t)]\nonumber\\
&=&\mathcal{P}[e^{-2\gamma t}]+\mathcal{P}[- e^{-2\gamma t}].\label{dis-core}
\end{eqnarray}
 From the Eq.(\ref{cla-core}) and (\ref{dis-core}) it is clear that $\mathcal{C}(\rho_{AB}(t))$ will always remain constant, while 
$ \mathcal{D}(\rho_{AB}(t))$ will decay from the beginning, which means there will be no such sudden transition in the 
evolution of classical correlations and discord. So, we can conclude that pure two-qubit entangled states will never show 
sudden transition.

To address our $2^{nd}$ question, we first consider two qubit Werner class of states, which is represented as: 
\begin{equation}
\rho_{AB} = \beta \lvert \psi^{-}\rangle\langle\psi^{-}\lvert + (1-\beta) \frac{\mathbf{1}_4}{4},
\end{equation}
where, $\beta$ represents the singlet fraction and $(1-\beta)$ represents random fraction. Simple calculations show that
for such states the  diagonal elements of the correlation tensor matrix are  all equal, i.e,
$\lvert c_1\lvert=\lvert c_2\lvert=\lvert c_3\lvert=\mbox{constant}= k(\mbox{say})$. Therefore, 
for such states $\mathcal{C}(\rho_{AB}(t))$ and $\mathcal{D}(\rho_{AB}(t))$ are found to be:
\begin{eqnarray}
\mathcal{C}(\rho_{AB}(t)) &=& \mathcal{I}(\sigma_ z ^A : \sigma_ z ^B) \nonumber \\
&=&\mathcal{P}[c_3]+\mathcal{P}[-c_3]\nonumber\\
&=&\mathcal{P}[k]+\mathcal{P}[-k],\\
\mathcal{D}(\rho_{AB}(t)) &=& \frac{1}{2}(\mathcal{P}[k+2k.e^{-2\gamma t}]\nonumber\\
&&+\mathcal{P}[k-2k.e^{-2\gamma t}]-2\mathcal{P}[k]). 
\end{eqnarray}
\begin{figure}[h!]
\centering
\includegraphics[height=6cm,width=6cm]{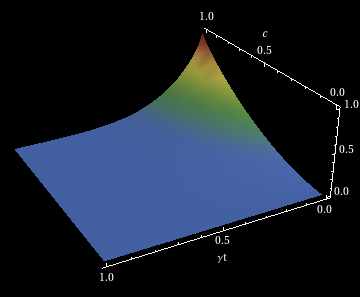}
\caption{(Color on-line) Evolution of $\mathcal{D}$ under phase-damping channel for Werner class of states. }\label{fig2}
\end{figure}
Hence, for phase damping noise classical correlations for this class of states will remain constant while the discord will
gradually decay without showing any kind of sudden transition. It is thus confirmed that mixedness is not the only factor
for sudden transition in classical and quantum decoherence. Along with mixed nature of the bipartite state $\rho_{AB}$,
it is also important that the state should have asymmetry in the correlations between the measurement outcomes of different
complementary observables.To say more specifically, $\mathcal{I}(\sigma_1^A:\sigma_1^B) >\mathcal{I}(\sigma_2^A:\sigma_2^B)=
\mathcal{I}(\sigma_3^A:\sigma_3^B).$

\section{Conclusions}\label{sec6}
We have shown that in case of certain Bell diagonal states, the sudden transition in the evolution of classical and 
quantum correlations under Markovian noise(phase damping channel) can be well understood in terms of $\emph{comlementary correlations}.$
For two qubit Bell-diagonal states $\sigma_x$ and $\sigma_z$ are complementary observables and for those specific 
Bell-diagonal states the overall complementary correlations are exactly equal to the mutual information($\mathcal{I}$) of the states.

We have also proved that two qubit pure entangled states will never show such type of sharp transition in the classical
and quantum decoherence. Taking the example of Werner class of states we have shown that the freezing property of discord 
($\mathcal{D}$), which is in contrast with entanglement $\emph{sudden death}$, is not inherent for all mixed entangled states
of Bell-diagonal class. Our analysis put forward an interesting question. Using complementary correlations, is it possible 
to find the general class of two qubit composite states exhibiting sudden change in the evolution of classical and 
quantum correlations? We hope that our findings will provide better insight to understand the evolution of classical
and quantum correlations of composite quantum systems, when subjected to noises.
\paragraph*{}
This work is funded by DST, Govt. of India. We are grateful to Dr. Guru Prasad Kar for useful discussions.


\begin{thebibliography}{99}
 \bibitem{Tele} C. H. Bennett, G. Brassard, C. Crépeau, R. Jozsa, A. Peres, and W. K. Wootters,
\href{https://journals.aps.org/prl/abstract/10.1103/PhysRevLett.70.1895}{Phys. Rev. Lett. {\bf 70}, 1895 (1993)}.

\bibitem{Dense} C. H. Bennett and S. J. Wiesner,
\href{https://journals.aps.org/prl/abstract/10.1103/PhysRevLett.69.2881}{Phys. Rev. Lett. {\bf 69}, 2881 (1992)}.

\bibitem{Schrodinger} E. Schrodinger,
\href{http://link.springer.com/article/10.1007\%2FBF01491891}{Naturwissenschaften {\bf 23}, 807 (1935)}.

\bibitem{Werner} R.F. Werner,
\href{http://journals.aps.org/pra/abstract/10.1103/PhysRevA.40.4277}{Phys. Rev. A {\bf 40}, 4277 (1989)}.

\bibitem{Horodecki} R. Horodecki, P. Horodecki, M. Horodecki, and  K. Horodecki, 
\href{http://journals.aps.org/rmp/abstract/10.1103/Rev Mod Phys.81.865}{Rev. Mod. Phys. {\bf 81}, 865 (2009)}.

\bibitem{SM} M. Horodecki, J. Oppenheim, A. Winter,
\href{http://www.nature.com/nature/journal/v436/n7051/full/nature03909.html}{Nature {\bf 436}, 673(2005)}

\bibitem{Zurek} H. Ollivier and W.H. Zurek,
\href{https://journals.aps.org/prl/abstract/10.1103/PhysRevLett.88.017901}{Phys. Rev. Lett. {\bf 88}, 017901 (2002)}.

\bibitem{Knill} E. Knill and R. Laflamme,
\href{https://journals.aps.org/prl/abstract/10.1103/PhysRevLett.81.5672}{Phys. Rev. Lett. {\bf 81}, 5672 (1998)}.

\bibitem{Streltsov} A. Streltsov, H. Kampermann, D. Bru$\ss{}$,
\href{https://journals.aps.org/prl/abstract/10.1103/PhysRevLett.106.160401}{Phys. Rev. Lett. {\bf 106}, 160401 (2011)}.

\bibitem{Dagmar} A. Streltsov, H. Kampermann, D. Bru$\ss{}$,
\href{https://journals.aps.org/prl/abstract/10.1103/PhysRevLett.108.250501}{Phys. Rev. Lett. {\bf 108}, 250501 (2012)}.

\bibitem{Cubitt} T.S. Cubitt, F. Verstraete, W. D\"{u}r, J.I. Cirac,
\href{https://journals.aps.org/prl/abstract/10.1103/PhysRevLett.91.037902}{Phys. Rev. Lett. {\bf 91}, 037902 (2003)}.

\bibitem{Chuan} T.K. Chuan, J. Maillard, K. Modi, T. Paterek, M. Paternostro, M. Piani
\href{https://journals.aps.org/prl/abstract/10.1103/PhysRevLett.109.070501}{Phys. Rev. Lett. {\bf 109}, 070501 (2012)}.

\bibitem{Datta} V. Madhok and A. Datta,
\href{https://journals.aps.org/pra/abstract/10.1103/PhysRevA.83.032323}{Phys. Rev. A {\bf 83}, 032323 (2011)}.

\bibitem{Adhikari} S. Adhikari and S. Banerjee
\href{https://journals.aps.org/pra/abstract/10.1103/PhysRevA.86.062313}{Phys. Rev. A {\bf 86}, 062313 (2012)}.

\bibitem{Groisman} B. Groisman, S. Popescu and A. Winter,
\href{https://journals.aps.org/pra/abstract/10.1103/PhysRevA.72.032317}{Phys. Rev. A {\bf 72}, 032317 (2005)}.

\bibitem{Schumacher} B. Schumacher and M.D. Westermoreland,
\href{https://journals.aps.org/pra/abstract/10.1103/PhysRevA.74.042305}{Phys. Rev. A {\bf 74}, 042305 (2006)}.

\bibitem{Vedral} L. Henderson and V. Vedral,
\href{https://iopscience.iop.org/0305-4470/34/35/315}{J. Phys. A: Math. Gen {\bf 34}, 06899 (2001)}.

\bibitem{Maziero1} J. Maziero, L. C C\'{e}leri, R. M. Serra and V. Vedral,
\href{https://journals.aps.org/pra/abstract/10.1103/PhysRevA.80.044102}{Phys. Rev. A {\bf 80}, 044102 (2009)}.

\bibitem{Maziero2} J. Maziero $\it {et~ al.}$,
\href{https://journals.aps.org/pra/abstract/10.1103/PhysRevA.81.022116}{Phys. Rev. A {\bf 81}, 022116 (2010)}.

\bibitem{Mazzola} L. Mazzola, J. Piilo and S. Maniscalco,
\href{https://journals.aps.org/prl/abstract/10.1103/PhysRevLett.104.200401}{Phys. Rev. Lett. {\bf 104}, 200401 (2010)}.

\bibitem{Sarandy} J. D. Montealegre, F. M. Paula, A. Saguia and M.S. Sarandy,
\href{https://journals.aps.org/pra/abstract/10.1103/PhysRevA.87.042115}{Phys. Rev. A {\bf 87}, 042115 (2013)}.

\bibitem{Fanchini} F. F. Fanchini $\it {et~ al.}$,
\href{https://journals.aps.org/pra/abstract/10.1103/PhysRevA.81.052318}{Phys. Rev. A {\bf 81}, 052107 (2010)}.

\bibitem{Yu1} T. Yu and J. H. Eberly,
\href{https://journals.aps.org/prl/abstract/10.1103/PhysRevLett.93.140404}{Phys. Rev. Lett. {\bf 93}, 140404 (2004)}.

\bibitem{Yu2} T. Yu and J. H. Eberly,
\href{https://journals.aps.org/prl/abstract/10.1103/PhysRevLett.97.140403}{Phys. Rev. Lett. {\bf 97}, 140403 (2006)}.

\bibitem{Lorenzo} L. Maccone,  D. Bru$\ss{}$ and C. Macchiavello,
\href{http://arxiv.org/abs/1408.6851}{arXiv:1408.6851}.

\bibitem{Wu} Shengjun Wu, Zhihao Ma, Zhihua Chen and Sixia Yu,
\href{http://nature.com/srep/2014/140207/srep04036/full/srep04036.html}{ Sci. Rep. {\bf 4}, 4036 (2014)}

\bibitem{chuang} Quantum Computation and Quantum Information, Michael. A Nielsen and Issac. L Chuang,

\bibitem{Fu} S. Luo and S. Fu
\href{https://journals.aps.org/pra/abstract/10.1103/PhysRevA.82.034302}{Phys. Rev. A {\bf 82}, 034302 (2010)}.

\bibitem{Dakic} B. Dakic, V. Vedral, C. Brukner,
\href{https://journals.aps.org/prl/abstract/10.1103/PhysRevLett.105.190502}{Phys. Rev. Lett. {\bf 105}, 190502 (2010)}.

\bibitem{Tufarelli} F. Ciccarello, T. Tufarelli, V. Giovannetti,
\href{https://iopscience.iop.org/1367-2630/16/1/013038}{New J. Phys. {\bf 16}, 013038 (2014)}.

\bibitem{Cerf} N. J. Cerf and C. Adami,
\href{https://journals.aps.org/prl/abstract/10.1103/PhysRevLett.79.5194}{Phys. Rev. Lett. {\bf 79}, 5194 (1997)}.

\bibitem{adesso} T.Tufarelli,T. MacLean, D. Girolami, R. Vasile, and G. Adesso,
\href{http://arxiv.org/abs/1301.3526}{arXiv:1301.3526}.

\bibitem{choi} M-D Choi, Lin. Alg. Appl. 10, 285–290 (1975)

\bibitem{Horodecki1} R. Horodecki and M. Horodecki,
\href{https://journals.aps.org/pra/abstract/10.1103/PhysRevA.54.1838}{Phys. Rev. A {\bf 54}, 1838 (1996)}.

\bibitem{zurek2} W.H Zurek, 
\href{http://journals.aps.org/rmp/abstract/10.1103/RevModPhys.75.715}{Rev. Mod. Phys {\bf 75}, 715 (2003)}.

\bibitem{wolf} M.M. Wolf and J.I. Cirac,
\href{http://arxiv.org/abs/0611057v3}{arXiv:0611057v3}.

\bibitem{kraus} K. Kraus, States, Effects and Operations: Fundamental Notions of Quantum Theory, Springer Verlag 1983





\end{thebibliography}
\end{document}